\documentclass[12pt,preprint]{aastex}

\shorttitle{The Twin Cepheids}
\shortauthors{D\'ek\'any et al.}

\begin{document}


\title{Discovery of a Pair of Classical Cepheids in an Invisible Cluster Beyond the Galactic Bulge}


\author{I.~D\'ek\'any\altaffilmark{1,2}, D.~Minniti\altaffilmark{3,1,7}, G.~Hajdu\altaffilmark{2,1}, J.~Alonso-Garc\'ia\altaffilmark{2,1}, M.~Hempel\altaffilmark{2}, T.~Palma\altaffilmark{1,2}, M.~Catelan\altaffilmark{2,1}, W.~Gieren\altaffilmark{1,4}, D.~Majaess\altaffilmark{5,6}}
\affil{$^1$Millennium Institute of Astrophysics, Santiago, Chile}
\affil{$^2$Instituto de Astrof\'isica, Facultad de F\'isica, Pontificia Universidad Cat\'olica de Chile, Av. Vicu\~na Mackenna 4860, Santiago, Chile}
\affil{$^3$Departamento de Ciencias F\'isicas, Universidad Andres Bello, Rep\'ublica 220, Santiago, Chile}
\affil{$^4$Departamento de Astronom\'ia, Universidad de Concepci\'on, Casilla 160-C, Concepci\'on, Chile}
\affil{$^5$Department of Astronomy \& Physics, Saint Mary's University, Halifax, NS B3H 3C3, Canada}
\affil{$^6$Mount Saint Vincent University, Halifax, NS B3M 2J6, Canada.}

\altaffiltext{7}{Also at Vatican Observatory, V-00120 Vatican City State, Italy}


\begin{abstract}
\noindent We report the discovery of a pair of extremely reddened classical Cepheid variable stars located in the Galactic plane behind the bulge, using near-infrared time-series photometry from the VVV Survey. This is the first time that such objects have ever been found in the opposite side of the Galactic plane. The Cepheids have almost identical periods, apparent brightnesses and colors. From the near-infrared Leavitt law, we determine their distances with $\sim1.5\%$ precision and $\sim8\%$ accuracy. We find that they have a same total extinction of $A(V)\simeq32$~mag, and are located at the same heliocentric distance of $\langle d\rangle = 11.4 \pm 0.9$~kpc, and less than $1$~pc from the true Galactic plane. Their similar periods indicate that the Cepheids are also coeval, with an age of $\sim48\pm3$~Myr, according to theoretical models. They are separated by an angular distance of only $18.3''$, corresponding to a projected separation of $\sim1$~ pc. Their position coincides with the expected location of the Far 3 kpc Arm behind the bulge. Such a tight pair of similar classical Cepheids indicates the presence of an underlying young open cluster, that is both hidden behind heavy extinction and disguised by the dense stellar field of the bulge. All our attempts to directly detect this ``invisible cluster'' have failed, and deeper observations are needed.
\end{abstract}





\section{Introduction}
\label{intro}

Classical Cepheids are ideal standard candles since their pulsation periods and high luminosities are tightly correlated \citep[i.e., the Leavitt Law,][]{1912HarCi.173....1L}. They are also pertinent proxies of young stellar populations, and have been employed to delineate the Milky Way's spiral arms \citep[e.g.,][and references therein]{2009MNRAS.398..263M}.

Classical Cepheids are by far the most valuable tracers of the Galactic spiral structure, since their period-luminosity (PL) relations are robust and accurately calibrated, and allow the intrinsic computation of the interstellar extinction. Furthermore, no putative assumptions of the  Galaxy's kinematics are required, and Cepheids can be used to map regions along the Galactic center and anticenter sightlines, where velocity-distance techniques fail. However, extreme extinction endemic to the Galactic disk has historically 
hampered the discovery of distant Cepheids at low Galactic latitudes in the 3rd and 4th Galactic quadrants. Pioneering near-infrared (NIR) variability surveys of the inner Galaxy, such as the IRSF Survey \citep{2011Natur.477..188M}, capable of tackling the problem of high differential extinction, have only recently started to probe into these inner Galactic regions.

The situation has been substantially changed by the VISTA Variables in the V\'ia L\'actea (VVV) Survey \citep{2010NewA...15..433M}. VVV is the only large NIR time-domain photometric survey of the bulge and the southern Galactic mid-plane. It has provided a $ZYJHK_s$ atlas of about a billion point sources, and has been acquiring time-series photometry in the $K_s$ band since 2010 \citep{2012A&A...537A.107S,2014Msngr.155...29H,2014arXiv1406.6727C}. VVV data present a historical opportunity for the census of distant classical Cepheids in the Galactic plane in the far side of the disk.

The VVV Galactic Cepheid Program (VGCP) is dedicated to discover and characterize distant Cepheids in the 3rd and 4th Galactic quadrants, concentrating on the detection young classical Cepheids lying close to the Galactic plane between $-2^{\circ}<b<2^{\circ}$ and $-65^{\circ}<l<10^{\circ}$. Due to the extreme reddening of $15\lesssim A(V)\lesssim50$, the Cepheids in this area are undetectable for all other existing and planned time-domain surveys.
In the framework of the VGCP, we expect to detect hundreds of classical Cepheids with distances up to the far edge of the Galactic disk, thus extending their census to the complete Southern Milky Way. These Cepheids will be used to map the spiral arm structure in the far side of the Milky Way's disk.
In addition to locating the spiral arms independently from other methods, they will also provide essential anchor points to velocity mapping surveys, supplying important constraints for the global kinematic properties of the Galaxy.

In this Letter, we report the first results of the VGCP: the discovery of an intriguing close pair of classical Cepheids located in the Galactic plane in a distant spiral arm beyond the bulge, that have almost identical physical parameters and distances.

\section{The Twin Cepheids}
\label{twincep}

In the framework of the VGCP, we are performing a massive and homogeneous photometric time-series analysis of point sources detected in the $K_s$-band VVV images. A complete in-depth analysis will be presented in a forthcoming paper, here we only give a brief summary. Our analysis is based on aperture photometry performed on single {\em detector frame stacks} (a.k.a. ``{\em pawprints}'') provided by the VISTA Data Flow System \citep[VDFS,][]{2004SPIE.5493..401E,2004SPIE.5493..411I} of the Cambridge Astronomy Survey Unit (CASU). A series of flux-corrected circular apertures are used, and optimal aperture sizes are determined for each object by minimizing the $\chi^2$ of residuals of the light-curve (LC) fits. The limiting $K_s$ magnitude is $\sim 16.5$~mag at small Galactic latitudes, which is sufficient for detecting Cepheids all the way through the Galactic plane up to the far side of the disk, with interstellar reddening up to $A(V)\sim50$~mag. Our analysis includes the positional cross-matching of individual CASU source catalogs, then a pre-selection of likely variable stars is performed on the assembled LCs based on various variability indices, and 
candidate variable objects ($\sim10\%$ of all sources) are then subjected to period analysis.

At the time of writing, we have analyzed the VVV $K_s$-band LCs of $\sim15$ million point sources towards the central bulge in low-latitude fields ($-2<b<2$ deg). The LCs typically have $50$ epochs over $4$ years, with additional $3$--$10$ epochs in the $J$ and $H$ bands. An initial search revealed a few hundred new distant Cepheid candidates in the $5$--$80$ day period range. A source of confusion in identifying these objects as classical Cepheids is brought in by the presence of foreground type II Cepheids, i.e., old, low-mass He-shell burning pulsating stars located in the bulge \citep[see, e.g.,][]{2011AcA....61..285S,2013MNRAS.429..385M}. They have similar periods and unlike in the optical, NIR LCs of Cepheids do not allow for a clear distinction between the two types. Therefore, in general, spectroscopic follow-up observations of the candidates are necessary for unambiguously verifying their nature, which is currently in progress by the VGCP team.

\begin{figure}
\begin{center}
\includegraphics[scale=1.0]{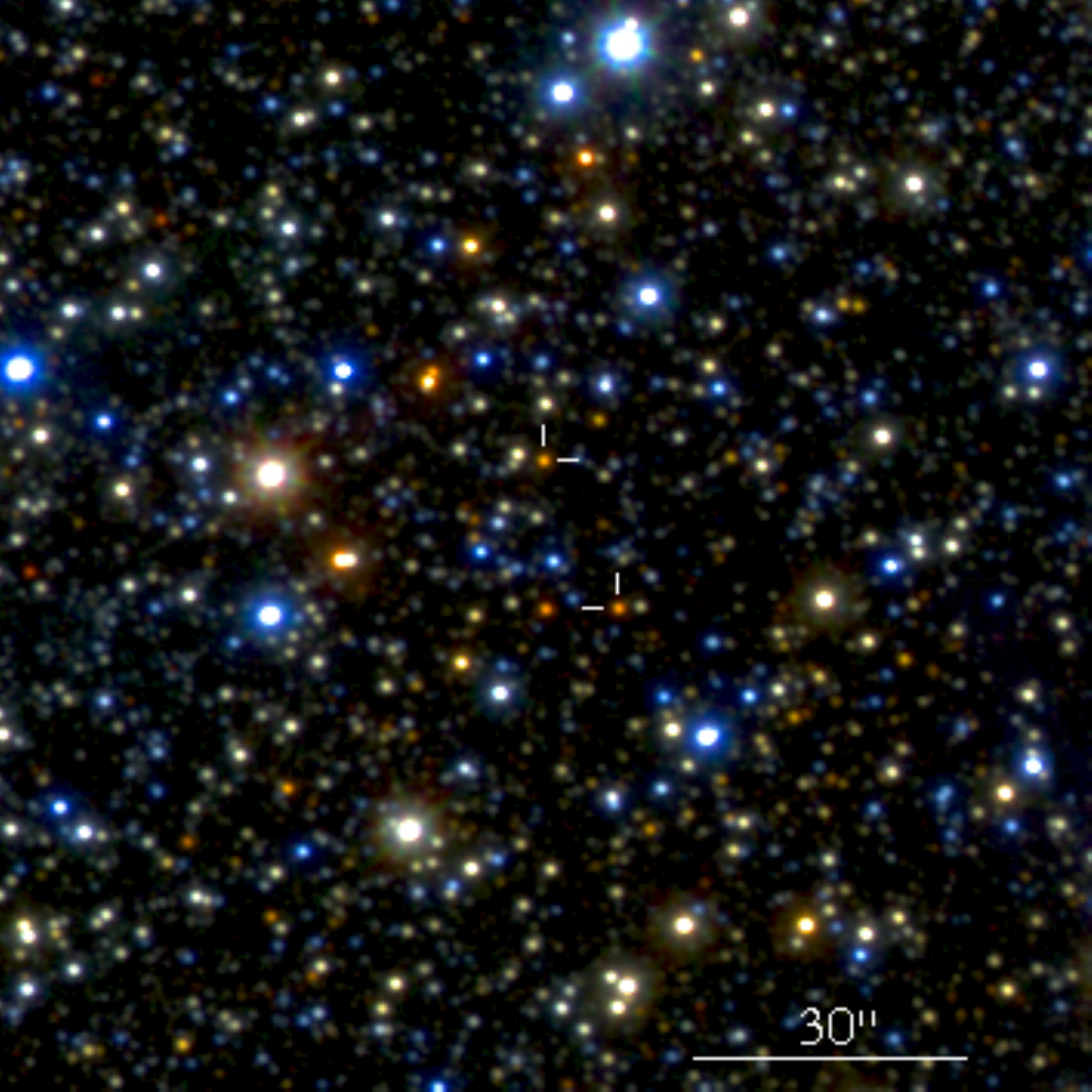}
\end{center}
\caption{False-color VVV deep stack image covering $1.7'\times1.7'$, showing the vicinity of the twin Cepheids (marked by white reticles). North is up, and East is to the left. The image is composed of $3$ $J$, $1$ $H$ and $24$ $K_s$ mosaic (a.k.a. \emph{tile}) images, used for the blue, green, and red channels, respectively.
}
\label{colorimg}
\end{figure}

Here we discuss an intriguing pair of objects discovered in our initial search for classical Cepheid candidates, that we call the Twin Cepheids. They are located at almost zero Galactic latitude, separated by an angular distance of only $18.3''$. The color finding chart of these two Cepheids is shown in Figure~\ref{colorimg}. Table~\ref{table} lists their positions and photometry, as well as the measured parameters such as periods, amplitudes, and their distances, that are discussed below.

\begin{figure}
\begin{center}
\includegraphics[scale=0.55]{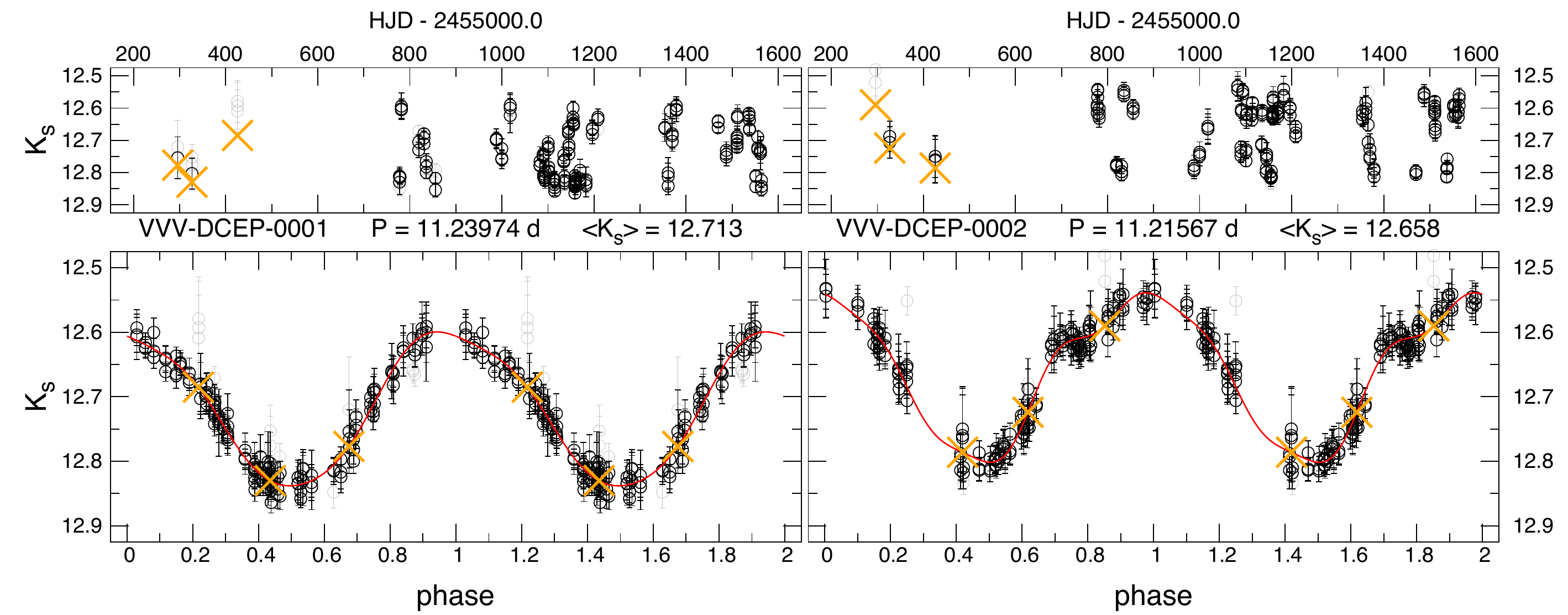}
\end{center}
\caption{VVV $K_s$-band light-curves and phase diagrams of the Twin Cepheids.
\emph{Red curves} denote the Fourier fits, \emph{orange crosses} show the predicted $K_s$ magnitudes at the times of the $H$-band measurements.}
\label{lcs}
\end{figure}

The LCs of the Twin Cepheids are shown\footnote{Photometric data are available online at the CDS; magnitudes are in the VISTA system.} in Fig.~\ref{lcs}. They have very similar periods and apparent magnitudes. We estimate that their ages are between 45 and 50 million years depending on their chemical compositions, according to the theoretical period-age relations of \cite{2005ApJ...621..966B}.

\begin{table}
\caption{Observed and computed parameters of the Twin Cepheids}
\label{table}
\begin{tabular}{cccccc}
\tableline\tableline
Star & $\alpha$ [hms]& $\delta$ [dms] & $l$ [deg]& $b$ [deg]& period [d] \\
\tableline
C1 & 18:01:24.49 & -22:54:44.6 & $6.99047$ & $0.00055$ & $11.23974$  \\
C2 & 18:01:25.09 & -22:54:28.3 & $6.99555$ & $0.00079$ & $11.21567$  \\
\end{tabular}
\begin{tabular}{cccccccc}
\tableline\tableline
Star & $\mathcal{A}_{tot.}\tablenotemark{\dag} (K_s)$ & $\langle K_s \rangle$ & $\langle H - K_s \rangle$ & $\langle J - K_s \rangle$ & E$(H-K_s)$ & $A(K_s)\tablenotemark{\ddag}$ & $d$ [pc] \\
\tableline
C1 & 0.24 & 12.71 (.02) & 2.05 (.05) & 6.42 (.20) & 2.00 & 3.27 & $11337 (951)$\\
C2 & 0.26 & 12.66 (.02) & 2.01 (.05) & 5.75 (.12) & 1.96 & 3.20 & $11397 (958)$\\
\tableline
\end{tabular}
\vskip2mm
Note: total errors are given in parentheses.
\tablenotetext{\dag}{Total amplitude of the $K_s$ light-curve.}
\tablenotetext{\ddag}{Assuming the \cite{2009ApJ...696.1407N} extinction curve.}
\end{table}

We determine their interstellar extinctions and distances with the same approach as in \cite{2013ApJ...776L..19D}, using PL relations in the $H$ and $K_s$ bands. We do not use the $J$-band magnitudes for this purpose because they are very close to the detection limit, and therefore have very large photometric errors. First, we compute the mean $\langle K_s\rangle$ magnitudes of the stars from the optimized Fourier fits of the LCs, and estimate their $\langle H-K_s\rangle$ mean color indices using the $K_s$ magnitudes at the phases of the $H$-band measurements derived from the LC fit. Then we compute the $M_H$ and $M_{K_s}$ mean absolute magnitudes from zero-point calibrated PL relations. From these quantities we calculate the color excess for each star by:
\begin{equation}
\label{ce}
E(H-K_s)=\langle H-K_s \rangle - (M_H - M_{K_s})~.
\end{equation}
An important advantage of using the $H$ and $K_s$ bands for this purpose is that variations in the $H-K_s$ color index of classical Cepheids through their pulsation cycles are negligible ($<0.05$~mag) independently of their periods \citep[see, e.g.,][]{2011ApJS..193...12M}. Moreover, the $\langle H-K_s\rangle$ of both stars are based on 3 $H$-band measurements taken at various different pulsation phases (see Fig.~\ref{lcs}), therefore any bias in the color excess derived from Eq.~(\ref{ce}) that is due to color variations ($\sigma(H-K_s)\simeq0.02$~mag for both stars) should be less than $0.01$~mag.

The absolute $A(K_s)$ extinction is determined by assuming the extinction curve found by \cite{2009ApJ...696.1407N}.
By transforming their absolute-to-selective extinction ratios from the 2MASS to the VISTA system using the equations given by CASU\footnote{\texttt{http://casu.ast.cam.ac.uk/surveys-projects/vista/technical/photometric-properties}}, we obtain:
\begin{equation}
\label{nishi}
A(K_{s})=1.63\times E(H - K_{s})~.
\end{equation}
Distances are computed by the following formula:
\begin{equation}
\label{dist}
\log R = 1 + 0.2\times (\langle K_s \rangle - A(K_{s}) - M_{K_s})~.
\end{equation}

The PL relations of classical and type II Cepheids are very different, and we use both in order to demonstrate that the Twin Cepheids must be of the first type. By assuming that they are type II, and employing the PL relations of \cite{2009MNRAS.397..933M}, we obtain a reddening value of $A(K_s)\simeq3.4$, and a distance of $R\simeq3.5$~kpc for both stars.
At this distance, the corresponding total reddening is unrealistically high when compared to the value of $A(K_s)=1.27$ obtained from the reddening map of \cite{2012A&A...543A..13G}, which corresponds to the mean cumulative extinction up to the mean distance of bulge red clump stars in the $2'$ angular radius of the Twin Cepheids. Even if there was a very strong and very localized anomaly in the foreground reddening, we should see that the bulge red clump stars at small angular separations from the Twin Cepheids are similarly reddened, which is clearly not the case (see Sect.~\ref{invoc}). Our conclusion is that both objects must be classical Cepheids.

To calculate the true distances of the Twin Cepheids, we employ the classical Cepheid PL relations of \cite{2012ApJ...759..146M}, by first transforming them from the 2MASS bands into the VISTA bands using the transformation equations provided by CASU, and then updating their zero-points to match the distance modulus of the Large Magellanic Cloud (LMC) derived by \cite{2013Natur.495...76P} using late-type eclipsing binaries. The resulting PL relations in the VISTA bands are as follows:
\begin{eqnarray}
\label{plr}
M_H &=& -3.228~[\pm 0.06] \times ( \log P - 1 ) -5.617~[\pm 0.048] \\
M_{K_s} &=& -3.269~[\pm 0.05] \times ( \log P - 1 ) -5.663~[\pm 0.048]~,
\end{eqnarray}
where the errors in the zero-points correspond to the total error in the distance modulus of the LMC. The resulting reddening values and distances are given in Table~\ref{table}. We note that the similarity of the reddenings under the type I and type II assumptions is caused by the very similar intrinsic colors of these objects.

We estimated the errors in the distances by Monte Carlo simulations, considering the errors in the $H$- and $K_s$-band VVV photometry and their zero-points, the mean magnitudes, the PL relations, the LMC distance modulus, and the \cite{2009ApJ...696.1407N} reddening law, by adding Gaussian random noise to these quantities with $\sigma$ made equal to their quoted errors. The resulting purely statistical errors are $150$~pc and $188$~pc for C1 and C2, respectively, which means an internal precision of $\sim1.5\%$. Indeed, the computed distances of the two stars differ by only $60$~pc. The systematic error is $\sim940$~pc for both stars, meaning a total accuracy of $8\%$ in the distance determination. We note that we might underestimate the systematic error in case the variations in the extinction curve along the Galactic plane exceed those found within the \cite{2009ApJ...696.1407N} fields.

Our conclusion is that the Twin Cepheids are located at a common distance of $\sim11.4$~kpc, and it would not be unexpected to find such young stars tracing a distant spiral arm.
Indeed, their positions coincide with the Far 3 Kpc Arm detected earlier in CO by \cite{2008ApJ...683L.143D}. At their distance, their angular separation corresponds to only $\sim 1$~pc. They also lie extremely close to the true Galactic plane, as expected for very young objects. In fact, by assuming a distance to the Galactic Center (GC) of $R_{\rm GC}=8.33$~kpc \citep{2013ApJ...776L..19D}, and that the Sun and the GC (i.e., Sgr A*) are located at $z_S=25$~pc and $z_{GC}=6$~pc off the IAU mid-plane, respectively \citep{2014arXiv1408.0001G}, then the IAU and the true Galactic planes cross at the same distance (to within the errors) where the Twin Cepheids are located, meaning that they are exactly in the true Galactic plane (i.e., less than $1$~pc from it).

\section{The Invisible Open Cluster}
\label{invoc}

The fact that the Twin Cepheids form a very close pair and share both the same extinction and distance suggests that this is not a projection effect, but they are indeed close to each other. This is supported by the fact that they are almost coeval, as indicated by their very similar periods (see Sect.~\ref{twincep}).
Since classical Cepheids are relatively rare objects, the probability of finding such a close pair with such similar properties just by chance is virtually zero. Therefore, we conclude that the Twin Cepheids indicate the presence of an underlying young open cluster in which they 
probably both still reside. Even in a cluster the probability of finding two such remarkably similar stars is small, but their membership in a heretofore undetected cluster appears to be a much more palatable explanation than chance alignment of two field stars.

In order to attempt a direct detection of this ``Invisible Cluster'', we analyzed the color-magnitude distribution of stars in the vicinity of the Twin Cepheids using a deep stack image from 24 best-seeing $K_s$-band mosaic images (a.k.a. \emph{tiles}) provided by CASU, using the SWarp software \citep{2002ASPC..281..228B}. For the $J$ band, we used a stack of 3 tiles, and one tile in $H$ because the rest of the images had poorer seeing (both Cepheids are beyond the limiting magnitude in $Z$ and $Y$). We performed PSF photometry using DoPhot \citep{1993PASP..105.1342S,2012AJ....143...70A} in a $8.5'\times8.5'$ field centered at $\alpha=$18:01:24.8, $\delta=$-22:54:36.4
(mid-point in between the Twin Cepheids), and calibrated it into the VISTA system using VDFS aperture photometry.

\begin{figure}
\includegraphics[angle=0,scale=0.8]{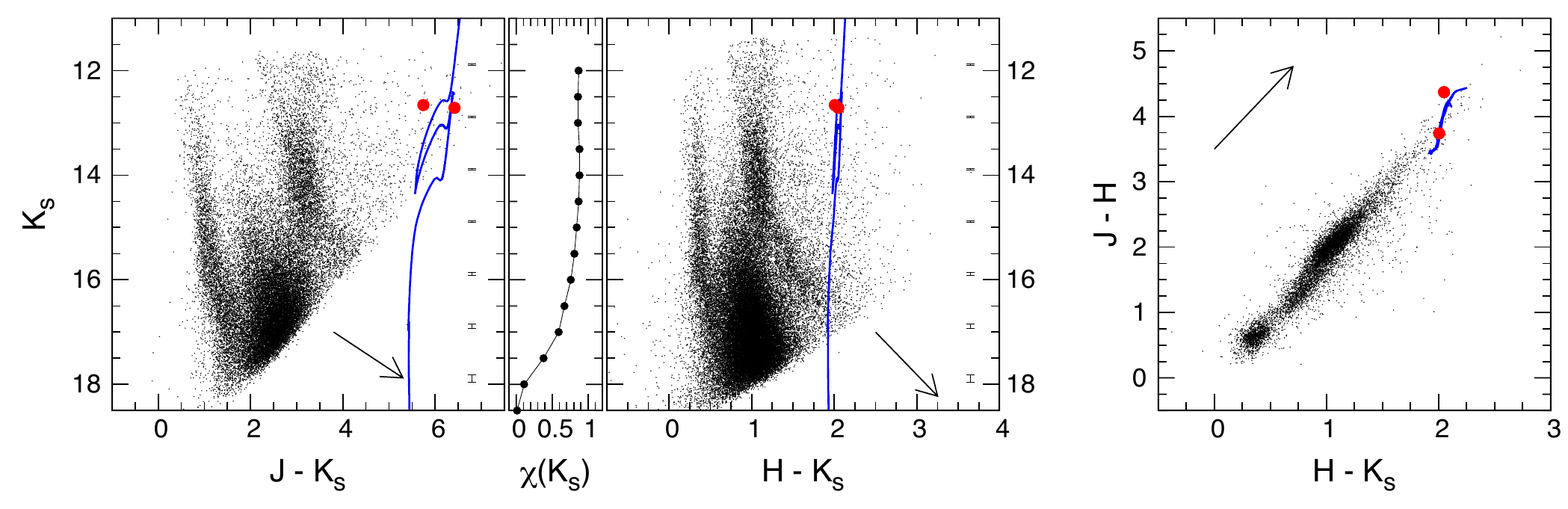}
\caption{Color-magnitude and color-color (for $K_s<15$\,mag) diagrams
of a $8.5'\times8.5'$ area around the Twin Cepheids.
Mean $K_s$ photometric errors are shown with error bars. The Cepheids' positions are marked with \emph{red points}, the arrows show the \cite{2009ApJ...696.1407N} reddening vectors. \emph{Blue lines} denote BaSTI isochrones
that best fit the properties of the Twin Cepheids (see text). The inset shows the $\chi$ completeness of $K_s$-band source detection.}
\label{cmds}
\end{figure}

Figure~\ref{cmds} presents the NIR CMDs and color-color diagrams of the field. The CMD is complex, but contains the expected prominent features: a local main sequence and a reddened bulge red giant branch, while the most numerous disk and bulge dwarf mix at fainter magnitudes (the unreddened bulge main sequence turnoff [MSTO] at a distance of $\sim8$~kpc should have $K_s\sim17.5$~mag). A distant red giant clump that appears extended along the reddening vector between $15\lesssim K_s \lesssim 17$, can also be observed in the $K_s~vs~H-K_s$ CMD. In addition, there are a number of very red point-sources with $H-K_s>1.5$ mag at all $K_s$ magnitudes, that are probably even more distant and reddened stars than the Twin Cepheids. We evaluated the completeness of source detection by adding a few thousand artificial stars to the $K_s$ image in every magnitude bin and measuring the ratio of those successfully extracted by DoPhot under the constraint of their magnitudes being recovered within $3\sigma$ (where $\sigma$ is the photometric error). The resulting $\chi (K_s)$ completeness curve is also shown in Figure~\ref{cmds}.

In order to identify the expected locus of stars in the Invisible Cluster, we compared our CMDs to the $50$-Myr Solar-composition canonical isochrone from the BaSTI stellar evolution database \citep{2004ApJ...612..168P}, best matching the expected ages of the Twin Cepheids. This isochrone was shifted to match the derived distances and reddenings of the Cepheids, and is shown as blue curves in Fig.~\ref{cmds}.

Neither the visual inspection of the CMDs, nor that of the deep stack images reveal any point-source overdensity at the expected location of the Invisible Cluster. We also employed the procedure by \cite{2008ApJ...686..279K} and
\cite{2009MNRAS.397.1748B} to search 
for any statistically significant overdensities. For this purpose, we used only sources selected in a vertical stripe around the isochrone on the $H~vs~H-K_s$ color-magnitude space using $1.6 \le (H-K_s) \le 2.8$. We detect a dense stellar structure at $\alpha=270.395~\deg$, $\delta=-22.897\deg$, but with a significance of only $5\%$, and both its position and significance change by small changes in the selection criteria. 


The failure of the direct detection of the Invisible Cluster can be expected if we consider that such a young cluster should have a very sparsely populated red giant branch (RGB). Since we are looking through the Galactic bulge and plane in the cluster's direction, field stars with various reddenings and distances contaminate its RGB, easily washing out any potential weak overdensity.
On the other hand, the cluster's MSTO, around which a higher density of cluster stars should be present, is expected to be at $K_s\simeq17$~mag, according to the BaSTI isochrone. The detection of an overdensity in our data at this magnitude level is also unlikely because the completeness of source detection is only $\sim50\%$ and quickly drops to a few per cent for fainter stars (see Fig.~\ref{cmds}), and the expected locus of the cluster MSTO overlaps with the abundant sequence of highly reddened background red clump stars, which can also easily wash out any weak signal in the overdensity detection test.

At the same time, we can identify a few very red sources that are good cluster star candidates (Fig.~\ref{colorimg}).
Finally, we note that because these two Cepheids and their companions within the Invisible Cluster are expected to share the rotation of the Galactic disk, their relative transversal motion velocity with respect to the Sun should be approximately $440$ km/s, corresponding to a proper motion of $\sim7.5$~mas/yr. We expect to measure this proper motion with the full VVV baseline in the next few years. 

\section{Conclusions}
\label{conclusions}

\begin{figure}
\begin{center}
\includegraphics[angle=0,scale=0.6]{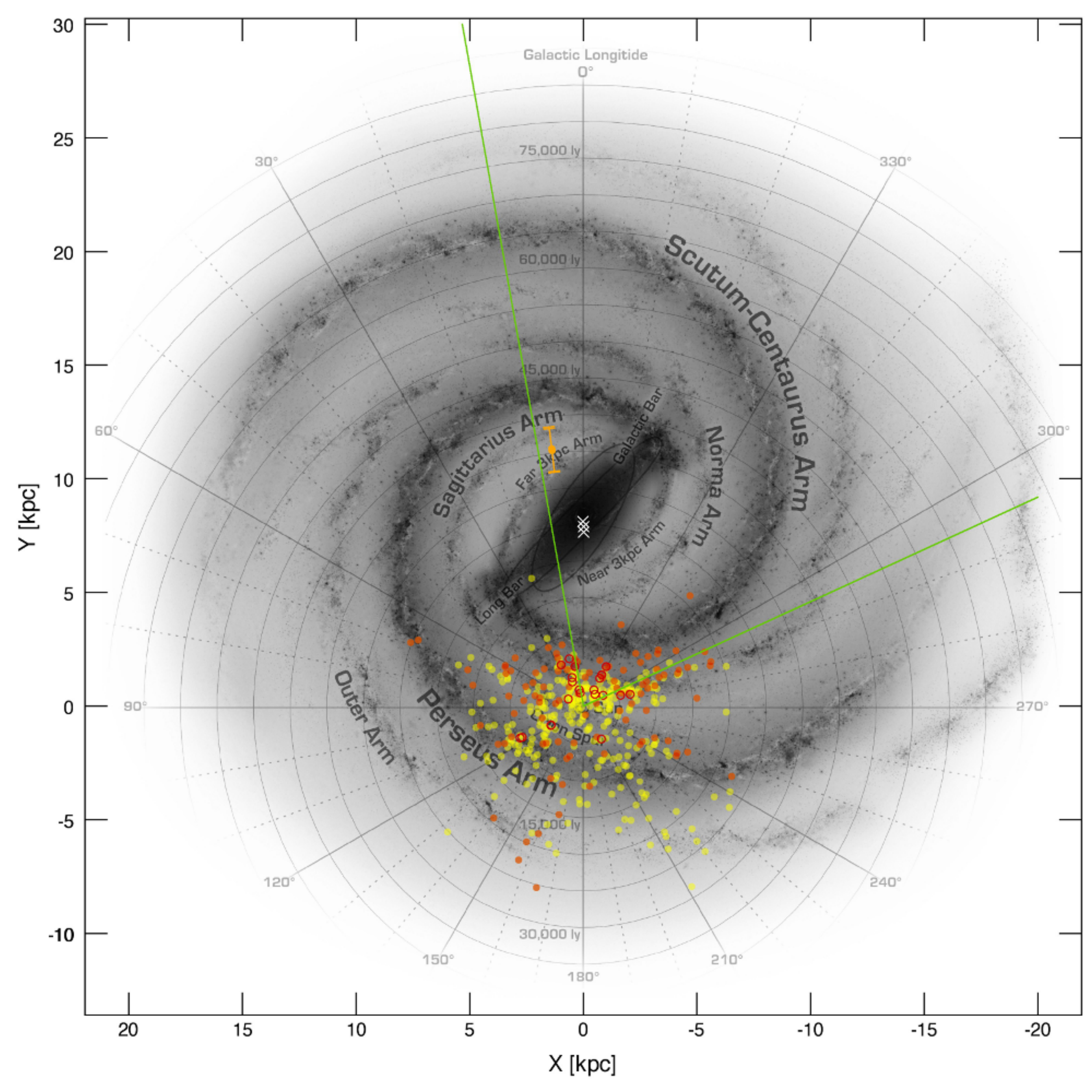}
\end{center}
\caption{Spatial distribution of known classical Cepheids from the DDO Cepheid database \citep{1995IBVS.4148....1F} projected onto the Galactic plane, overlaid on a sketch of the Milky Way by Robert Hurt. \emph{Yellow} and \emph{red points} show objects with periods shorter and longer than $10$ days, respectively, Cepheids in open clusters \citep{2013MNRAS.434.2238A} are marked with \emph{red circles}. \emph{White crosses} show the known classical Cepheids in the nuclear bulge \citep{2011Natur.477..188M}. The positions of the Twin Cepheids with their errors are indicated with orange points. The green lines denote the Galactic latitudinal coverage of the VVV Survey.}
\label{map}
\end{figure}

As part of the ongoing VVV Galactic Cepheid Program, we have presented the discovery of two heavily reddened ($A_V\simeq32$ mag) and distant classical Cepheids with almost identical properties.
Using NIR PL relations, we find that they are located at the same distance of $\langle d\rangle=11.37\pm0.17\pm0.94$~kpc, in the Far 3 Kpc Arm beyond the bulge. Their similarity and closeness suggest that they belong to the same young open cluster at ($l,b$)= 6.9930, +0.0005 deg, that is disguised by source crowding and extinction. The discovery reported here is both the first detection of classical Cepheids in the far side of the Galactic plane, and the first (indirect) detection of an open cluster in the observational shadow of the bulge. Our results also provide independent confirmation for the Far 3 Kpc Arm. The current census of Galactic classical Cepheids is illustrated by Figure~\ref{map}. It has been historically confined to the close side of the Galaxy, due to high interstellar extinction along the Galactic plane. 
The only confirmed classical Cepheids beyond the bulge were reported by \cite{2014Natur.509..342F}, but they are far from the Galactic plane, in the flared outer disk. The VVV Survey is now enabling us to pierce through tens of magnitudes of optical interstellar extinction and extend the census of Cepheids to the far side of the disk, with the ultimate goal of completing the face-on map of our Galaxy. 

The Twin Cepheids also provide an important proof of concept for the VGCP, by demonstrating its capabilities of detecting distant and extremely reddened Cepheids and measuring their distances with an excellent internal precision, and an accuracy better than $10\%$ (dominated by the uncertainty in the extinction law). Assuming the discovery of a few hundred classical Cepheids by the VGCP, we expect that this accuracy will be sufficient for distinguishing between distant spiral arms up to the far edge of the disk, and also trace possible satellite galaxies that lie beyond it. Mining these VVV Cepheids warrants interesting results.

\acknowledgments

We gratefully acknowledge the use of data from the ESO Public Survey program 179.B-2002 taken with the VISTA telescope, and data products from the Cambridge Astronomical Survey Unit. Funding from the BASAL CATA through grant PFB-06, the Chilean Ministry of Economy through ICM grant P07-021-F, and from FONDECYT through projects 1141141, 1130196 and 3130552 are also acknowledged.

{\it Facilities:} \facility{VISTA}.

\clearpage

\end{document}